\documentclass[aps, prd, a4paper, 12pt, amsmath, preprintnumbers, nofootinbib, tightenlines, longbibliography, notitlepage]{revtex4-1}\pdfoutput=1
\usepackage[english]{babel}
\usepackage[utf8]{inputenc} 
\usepackage{bbm}
\usepackage{graphicx}
\usepackage{color}
\usepackage{hyperref}

\graphicspath{{pics/}}

\newcommand{\bhh}[1]{\big(#1\big)}

\newcommand{\abs}[1]{\lvert#1\rvert}

\newcommand{\md}{\mathrm{d}}

\DeclareMathOperator{\pathord}{\mathcal{P}}

\DeclareMathOperator{\bigO}{\mathcal{O}}

\hypersetup{
	pdfauthor={Maarten van de Meent},
	pdftitle={The Geometry of Massless Cosmic Strings},
	pdfkeywords={ Massless cosmic strings - Exact solution - Holonomy - Classical general relativity}}

\begin{document}
\preprint{ITP-UU-12/44}
\preprint{SPIN-12/41}
\title{The Geometry of massless cosmic strings}

\author{Maarten \surname{van de Meent}}
\affiliation{Institute for Theoretical Physics and Spinoza Institute,\\ Utrecht University,\\ P.O. Box 80.195, 3508 TD Utrecht, the Netherlands}%
\email{M.vandeMeent@uu.nl}

\date{\today}
\begin{abstract}
We study the geometry generated by a massless cosmic string. We find that this is given by a Riemann flat spacetime with a conical singularity along the worldsheet of the string. The geometry of such a spacetime is completely fixed by the holonomy of a simple loop wrapping the conical singularity. In the case of a massless cosmic string, this holonomy is a null-rotation/parabolic Lorentz transformation with a parabolic angle given by the linear energy density of the cosmic string. This description explicitly shows that there is no gravitational shockwave accompanying the massless cosmic string as has been suggested in the past. To illustrate the non-singular nature of the surrounding geometry, we construct a metric for the massless cosmic string that is smooth everywhere outside the conical singularity.
\end{abstract}
\maketitle

\section{Introduction}
Cosmic strings as introduced by Kibble in 1976 \cite{Kibble:1976sj} have been extensively studied in the 1980s and 1990s \cite{Hindmarsh:1994re, vilenkin2000cosmic}. One of their main attractions was that they provided an alternative to inflation as a source for primordial density fluctuations. However, these predictions turned out to be incompatible with the precision measurements of the CMB \cite{Bevis:2004wk}, leading to a loss of interest in the subject.  

Nonetheless, some interest in the subject has remained. One reason is that studies have shown that the generation of cosmic strings appears to be a generic feature of GUT phase transitions \cite{Jeannerot:2003qv}. Moreover, string inflation models may lead to the production of cosmic scale superstrings (see \cite{Polchinski:2004ia} for a review).

In traditional studies, cosmic strings are assumed to have a non-zero mass density. Consequently, the geometry of their gravitational field can be analysed in their rest frame. One of the interesting features of cosmic strings is that this geometry is flat everywhere except for a conical singularity along the string.

In 2008, 't Hooft introduced a locally finite model for gravity in 3+1 dimensions, in which the fundamental excitations are straight cosmic strings, leading to a piecewise flat geometry \cite{hooft:2008,hooft:2009,meent:2010,PHDthesis}. Besides the usual massive modes, this model also allows massless cosmic strings. In fact, the massless strings turn out to be a crucial ingredient to recover dynamical gravity in the large scale limit of the model \cite{meent:2010b,meent:2011}.

Compared to their massive cousins, massless cosmic strings have received very little attention in the literature. The treatments that do exist \cite{Barrabes:2002hn, Lousto:1990wn}, are very brief and, at best, incomplete. In this paper, we will present a complete geometric description of massless cosmic strings.

In section \ref{sec:statstring} we review the geometry of a stationary massive cosmic string. In section \ref{sec:movstring}, we show how this can be boosted to give the geometry of a massive string moving at an arbitrary (subluminal) velocity.

In section \ref{sec:ASstring}, we follow the procedure of Aichelburg and Sexl \cite{Aichelburg:1970dh} to take the ultra-relativistic limit of the massive string, and obtain the metric of a massless cosmic string moving at the speed of light. Like the Aichelburg--Sexl metric for a massless particle, this metric has a $\delta$-singularity along a null-plane travelling with the string. However, this feature is rather misleading since it is not accompanied by a singularity in the Riemann curvature. From this metric, we obtain the holonomy of a single loop around the cosmic string. Since the spacetime is flat everywhere except along the string, this gives us the holonomy of \emph{any} loop. Consequently, this holonomy encodes all the geometric information.

Instead of an ordinary rotation, the holonomy of a massless string is a null-rotation, i.e. a Lorentz transformation that leaves a null-vector invariant. In section \ref{sec:cutpaste}, we use this holonomy to realize the geometry of a massless string as a conical defect using a null-rotation rather than a normal rotation.

In section \ref{sec:smooth}, we resolve the $\delta$-singularity pathologies of the metric found in section  \ref{sec:ASstring}. From the holonomy of the massless cosmic string we construct a metric that is smooth everywhere except along the string itself.

\section{Stationary Cosmic Strings}\label{sec:statstring}
Conventionally, the metric for a stationary cosmic string of infinitely length oriented along the $z$-axis is written as,
\begin{equation}\label{eq:massivestring}
\md s^2 = -\md t^2 + \md r^2 + (1-\frac{\mu}{2\pi})^2 r^2\md\theta^2 +\md z^2.
\end{equation}
The Riemann curvature of this metric vanishes everywhere except along the $tz$-hyperplane (the string worldsheet). The corresponding energy--momentum tensor can easily be calculated by thickening the string (e.g. see chapter 2.1 of \cite{PHDthesis}). If we take $\mu$ to be a function of $r$, the above metric corresponds to a matter source with energy--momentum,\footnote{Note that we are using units where $c=\hbar=8\pi G =1$.}
\begin{equation}
T = \frac{1}{2\pi\sqrt{-g}}(r \mu''(r)+2\mu'(r))(\md t^2-\md z^2).
\end{equation}
Assuming that $\mu(0)=0$ and $\mu'(R)=0$, the total energy--momentum tensor contained in a cylinder of radius $R$ and length $L$ is therefore,
\begin{equation}
L \mu(R) (\md t^2-\md z^2).
\end{equation}
In the case that $\mu$ is constant, the metric \eqref{eq:massivestring}  consequently can be viewed as describing an infinitely thin string with a mass per unit length of $\mu$. 

The holonomy of an arbitrary loop $\gamma(\lambda)$ around the cosmic string is given by
\begin{equation}\label{eq:holonomy}
(P_\gamma)_{\phantom{\mu}\nu}^\mu =\pathord \exp\bhh{-\int_\gamma \Gamma^\mu_{\sigma\nu} \frac{\md \gamma^\sigma}{\md \lambda} \md \lambda},
\end{equation}
where the $\pathord$ indicates that the matrix exponential is path-ordered and $\Gamma^\mu_{\sigma\nu}$ are the Christoffel symbols of the metric. Since, the geometry outside the cosmic string is Riemann flat, the holonomy can only depend on the winding number of the loop around the cosmic string, and we can thus speak of \emph{the} holonomy of the cosmic string. It also suffices to calculate the holonomy of a simple loop with $t=z=0$, $r=(1-\mu/2\pi)^{-1}$ constant, and $\theta=2\pi\lambda$. The integrand above then becomes
\begin{equation}
-\Gamma^\mu_{\sigma\nu} \frac{\md \gamma^\sigma}{\md \lambda} = \begin{pmatrix}
0 & 0 & 0 & 0\\
0 & 0 & 2\pi -\mu & 0\\
0 & \mu - 2\pi & 0 & 0\\
0 & 0 & 0 & 0
\end{pmatrix}.
\end{equation}
Because this is independent of $\lambda$, the path ordering in \eqref{eq:holonomy} becomes trivial, and the matrix exponential can be computed directly yielding,
\begin{equation}\label{eq:Hsstring}
(P_\gamma)_{\phantom{\mu}\nu}^\mu = \begin{pmatrix}
1 & 0 & 0 & 0\\
0 & \cos\mu & -\sin\mu & 0\\
0 & \sin\mu & \cos\mu & 0\\
0 & 0 & 0 & 1
\end{pmatrix}.
\end{equation}
This identifies the geometry of the massive stationary string \eqref{eq:massivestring} as a conical singularity with a deficit angle equal to the mass density per unit length $\mu$. This holonomy encodes all there is to know about the orientation and mass of the cosmic string. The only missing piece of information is the location of the cosmic string, which can be included also by calculating the Poincaré holonomy \cite{hooft:2008}.

\section{Moving strings}\label{sec:movstring}

The metric for a moving  string can be found by boosting metric \eqref{eq:massivestring}. To do this we first introduce ``Cartesian'' coordinates $x = r \cos\theta$ and $y = r\sin\theta$, in which the metric becomes
\begin{equation}\label{eq:massivestring2}
\md s^2 = -\md t^2 + \md x^2 + \md y^2 +\md z^2 -
\frac{\mu}{2\pi}(2-\frac{\mu}{2\pi})\frac{(y\md x - x\md y)^2}{x^2+y^2}.
\end{equation}
Subsequently, we introduce new ``boosted'' (lightcone) coordinates,
\begin{alignat}{2}
\bar{t} &= t\cosh\chi  + x\sinh\chi &= \frac{v-u}{\sqrt{2}},\\
\bar{x} &= x\cosh\chi  + t\sinh\chi &= \frac{u+v}{\sqrt{2}}.
\end{alignat}
In these coordinates metric \eqref{eq:massivestring2} becomes
\begin{equation}\label{eq:movingstring}
\md s^2 =2\md u\md v + \md y^2 +\md z^2 -
\frac{\mu}{2\pi}(2-\frac{\mu}{2\pi})\frac{\bhh{e^{\chi}(y\md u - u\md y)+e^{-\chi}(y\md v - v\md y)}^2}{e^{2\chi}u^2+e^{-2\chi}v^2+2(uv+y^2)}.
\end{equation}
Since metric \eqref{eq:massivestring2} describes a cosmic string standing still at the origin, the new metric \eqref{eq:movingstring} describes a cosmic string moving in the positive $\bar x$ direction with velocity $V=\tanh\chi$.

Metric \eqref{eq:massivestring2} corresponded to an energy--momentum source,
\begin{equation}
 \frac{\mu}{1-\frac{\mu}{2\pi}}\delta(x)\delta(y)(\md t^2-\md z^2).
\end{equation}
Consequently, the boosted cosmic string corresponds to a source,
\begin{equation}\label{eq:Tmovingstring}
 \frac{\mu}{(1-\frac{\mu}{2\pi})\cosh\chi} \delta(\bar{x}- \bar{t} \tanh\chi)\delta(y)\bhh{(\cosh\chi\md \bar{t} - \sinh\chi\md \bar{x})^2-\md z^2}.
\end{equation}

\section{The Aichelburg--Sexl boosted string}\label{sec:ASstring}
The metric for a massless string travelling at the speed of light can be found using the technique introduced by Aichelburg and Sexl \cite{Aichelburg:1970dh} to derive the metric of a massless point particle (see for example \cite{Barrabes:2002hn}). The idea is to take the boosted metric for a moving string and increase the rapidity $\chi$, while keeping the energy density as measured in the laboratory frame constant.

The latter is found from equation \eqref{eq:Tmovingstring} as
\begin{equation}
\iint T_{\bar{t}\bar{t}}\sqrt{g^{(2)}}\md x\md y = \mu\cosh\chi,
\end{equation}
where $g^{(2)}$ is the determinant of the metric induced on the $xy$-plane. Consequently, setting $\mu= \bar\mu/\cosh\chi$ keeps the observed energy density constant.

For large values of $\chi$, the $\chi$ dependent part of metric \eqref{eq:movingstring} becomes
\begin{equation}
-\frac{2\bar\mu}{\pi} \frac{e^{-\chi}(y\md u - u\md y)^2}{u^2 +e^{-4\chi}v^2+2e^{-2\chi}(u v+y^2)}+\bigO(e^{-2\chi}).
\end{equation}
The pointwise limit of this term as $\chi\to\infty$ is zero. However, using the identity,
\begin{equation}
\lim_{b\to 0} \frac{b}{u^2+b^2} = \pi\delta(u),
\end{equation}
it can be shown \cite{Barrabes:2002hn} that in a distributional sense
\begin{equation}
\lim_{\chi\to\infty} \frac{e^{-\chi}}{u^2 +e^{-4\chi}v^2+2e^{-2\chi}(u v+y^2)} = \frac{\pi}{\sqrt{2}\abs{y}}\delta(u).
\end{equation}
Consequently, metric \eqref{eq:movingstring} becomes
\begin{equation}\label{eq:masslessstring}
\md s^2 =2\md u\md v + \md y^2 +\md z^2 -
\sqrt{2}\bar\mu\abs{y}\delta(u)\md u^2,
\end{equation}
in the Aichelburg--Sexl limit. The extent to which general relativity with distribution valued tensors makes sense has been discussed at length in the literature (see \cite{Steinbauer:2006qi} for a review). The upshot of these discussions, as far as this article is concerned, is that as long as physical quantities such as the curvature calculated from the metric do not contain pathological products of distributions, one can proceed normally. Doing so, it is straightforward to calculate its Riemann curvature\footnote{With an expression like $(\md u\wedge\md y)^2$ we mean the symmetrized tensor product of the wedge products. }
\begin{equation}\label{eq:Rmstring}
R = -2\sqrt{2} \bar\mu\delta(u)\delta(y) (\md u\wedge\md y)^2,
\end{equation}
and the Einstein curvature
\begin{align}
G &= \sqrt{2}\bar\mu\delta(u)\delta(y) \md u^2\\
&=\bar\mu \delta(x-t)\delta(y) (\md t-\md x)^2.\label{eq:EMmstring}
\end{align}
We thus see that the Aichelburg--Sexl boosted metric \eqref{eq:masslessstring} indeed describes a stringlike curvature defect on the lightlike surface $x-t=y=0$. We also confirm that the linear energy density as measured in the laboratory frame is indeed $\bar\mu$. Equation \eqref{eq:Rmstring} also reveals a rather unsatisfactory property of metric \eqref{eq:masslessstring}: while the metric has a $\delta$-singularity on the entire $u=0$ hypersurface, its curvature vanishes almost everywhere on that surface, except along the $y=0$ locus. In the past the presence of a $\delta$-singularity in \eqref{eq:masslessstring} has led to the mistaken conclusion that a gravitational shockwave is present (e.g. implicitly in \cite{Lousto:1990wn}). In section \ref{sec:smooth}, we will construct a continuous metric that is smooth everywhere except along the singularity.

Since  metric \eqref{eq:masslessstring} is flat everywhere outside the singular surface $x-t=y=0$, it describes a conical singularity with a well-defined holonomy. As was the case with the stationary string, the holonomy completely captures the geometric data of the massless string spacetime  \eqref{eq:masslessstring}. We can find this holonomy either by calculating the Aichelburg--Sexl boost of the holonomy of the stationary string \eqref{eq:Hsstring} or directly from metric \eqref{eq:masslessstring}. We will do both as a consistency check.

By Lorentz covariance the holonomy of a moving cosmic string is given by
\begin{equation}\label{eq:Hmvstring}
Q_{\phantom{\mu}\nu}^\mu(\chi) = \begin{pmatrix}
1 + 2\sin^2(\frac{\mu}{2})\sinh^2\chi
	& -\sin^2(\frac{\mu}{2}) \sinh 2\chi
		& -\sin\mu\sinh\chi 
			& 0\\
 \sin^2(\frac{\mu}{2}) \sinh 2\chi
 	& 1 - 2\sin^2(\frac{\mu}{2})\cosh^2\chi
		& -\sin\mu\cosh\chi 
			& 0\\
-\sin\mu\sinh\chi 
	& \sin\mu\cosh\chi 
		& 1-2\sin^2(\frac{\mu}{2})
			& 0\\
0 & 0 & 0 & 1
\end{pmatrix}.
\end{equation}
For large enough values of $\chi$,
\begin{equation}
\sin\mu = \frac{1}{2}\bar\mu e^{-\chi} + \bigO(e^{-2\chi}).
\end{equation}
Consequently, the Aichelburg--Sexl limit of the holonomy is
\begin{equation}\label{eq:Hmstring}
\lim_{\chi\to\infty}Q_{\phantom{\mu}\nu}^\mu(\chi) = \begin{pmatrix}
1 + \frac{\bar\mu^2}{2}
	& -\frac{\bar\mu^2}{2}
		& -\bar\mu
			& 0\\
 \frac{\bar\mu^2}{2}
 	& 1 - \frac{\bar\mu^2}{2}
		& -\bar\mu
			& 0\\
-\bar\mu
	& \bar\mu
		& 1
			& 0\\
0 & 0 & 0 & 1
\end{pmatrix},
\end{equation}
or in lightcone coordinates $x^\mu = \{u,v,y,z\}$,
\begin{equation}\label{eq:Hmstringlc}
\lim_{\chi\to\infty}Q_{\phantom{\mu}\nu}^\mu(\chi) = \begin{pmatrix}
1 
	& 0
		& 0
			& 0\\
-\bar\mu^2
 	& 1 
		& -\sqrt{2}\bar\mu
			& 0\\
\sqrt{2}\bar\mu
	& 0
		& 1
			& 0\\
0 & 0 & 0 & 1
\end{pmatrix}.
\end{equation}
Lorentz transformations of the form \eqref{eq:Hmstring} and \eqref{eq:Hmstringlc} are known as \emph{null-rotations} or \emph{parabolic Lorentz transformations}. They are characterized by the fact that they leave a null-vector (the $v$-axis in this case) invariant, or alternatively by the fact that their generators are the nilpotent matrices in the Lorentz Lie algebra. In particular, the parameter $\bar\mu$ (called the \emph{parabolic angle}) is additive under multiplication of the matrices.

\begin{figure}[t]
\centering
\def\svgwidth{0.5\columnwidth}
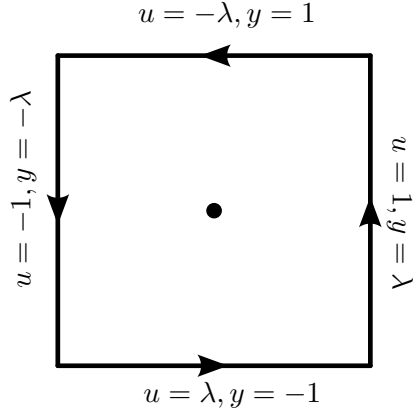
\caption{Specification of a loop wrapping the massless cosmic string (dot in the center) used to calculate the holonomy, drawn in the $uy$-plane.}\label{fig:hloop}
\end{figure}

As a check of the limiting procedure we also calculate the holonomy directly from  metric \eqref{eq:masslessstring}. We choose a square loop $\gamma$ around the lightlike singularity as indicated in figure \ref{fig:hloop}. The loop consists of four sides
\begin{align}
\gamma_1^\mu(\lambda) &= (\lambda,0,-1,0)\\
\gamma_2^\mu(\lambda) &= (1,0,\lambda,0)\\
\gamma_3^\mu(\lambda) &= (-\lambda,0,1,0)\\
\gamma_4^\mu(\lambda) &= (-1,0,-\lambda,0),
\end{align}
with $\lambda$ running from $-1$ to $1$ on each side.

For sides $2$ and $4$, $\Gamma^{\mu}_{\sigma\nu}\tfrac{\md \gamma_{2,4}^\sigma}{\md \lambda}$ vanishes, so they do not contribute to the holonomy. For sides $1$ and $3$,
\begin{equation}
-\Gamma^{\mu}_{\sigma\nu}\frac{\md \gamma_{1,3}^\sigma}{\md \lambda}=\begin{pmatrix}
0 & 0 & 0 & 0\\
-\frac{\bar\mu}{\sqrt{2}}\delta'(\lambda) & 0 & -\frac{\bar\mu}{\sqrt{2}}\delta(\lambda)  & 0\\
\frac{\bar\mu}{\sqrt{2}}\delta(\lambda) &0 & 0 & 0\\
0 & 0 & 0 & 0
\end{pmatrix},
\end{equation}
which commute for different values of $\lambda$ making the path-ordering in equation \eqref{eq:holonomy} trivial. Consequently, the holonomy of the entire loop $\gamma$ is given by
\begin{equation}
(P_\gamma)_{\phantom{\mu}\nu}^\mu =\begin{pmatrix}
1 & 0 & 0 & 0\\
-\frac{\bar\mu^2}{4} & 1 & -\frac{\bar\mu}{\sqrt{2}}  & 0\\
\frac{\bar\mu}{\sqrt{2}} &0 & 1 & 0\\
0 & 0 & 0 & 1
\end{pmatrix}^2 =\begin{pmatrix}
1 & 0 & 0 & 0\\
-\bar\mu^2 & 1 & -\sqrt{2}\bar\mu  & 0\\
\sqrt{2}\bar\mu &0 & 1 & 0\\
0 & 0 & 0 & 1
\end{pmatrix},
\end{equation}
which matches the holonomy \eqref{eq:Hmstringlc} obtained through the Aichelburg--Sexl limiting procedure.

The authors of \cite{Barrabes:2002hn} identify $\bar\mu$ as the deficit angle  of the conical singularity as measured in the laboratory frame. To us, this interpretation seems physically unacceptable. In particular, it would imply that something special should happen for an observer in the laboratory frame at the value $\bar\mu=2\pi$, since a deficit angle of $2\pi$ would imply a collapse of the spacial slice. However, neiter the metric \eqref{eq:masslessstring} nor the holonomy \eqref{eq:Hmstringlc} give us any reason to believe that there is anything special about the value $\bar\mu=2\pi$.

At least part of the problem is that equal time slices in \eqref{eq:masslessstring} do not correspond to anything that an observer would recognize as a spacial slice. This is because any geodesic crossing the $u=0$ hypersurface makes a jump $\Delta v = \bar\mu\abs{y}/\sqrt{2}$. Effectively, the metric \eqref{eq:masslessstring} describes two independent coordinate patches ($u<0$ and $u>0$), which are identified along $u=0$ in an awkward way.

The holonomy calculated above, gives a much cleaner geometrical interpretation of the parameter $\bar\mu$, namely as the parabolic angle corresponding to the holonomy of the defect. In the next section, we will find the deficit angle of the singularity as measured on an equal time slice.

\section{Cut-and-paste geometry}\label{sec:cutpaste}
\begin{figure}[t]
\centering
\includegraphics[width=0.5\textwidth]{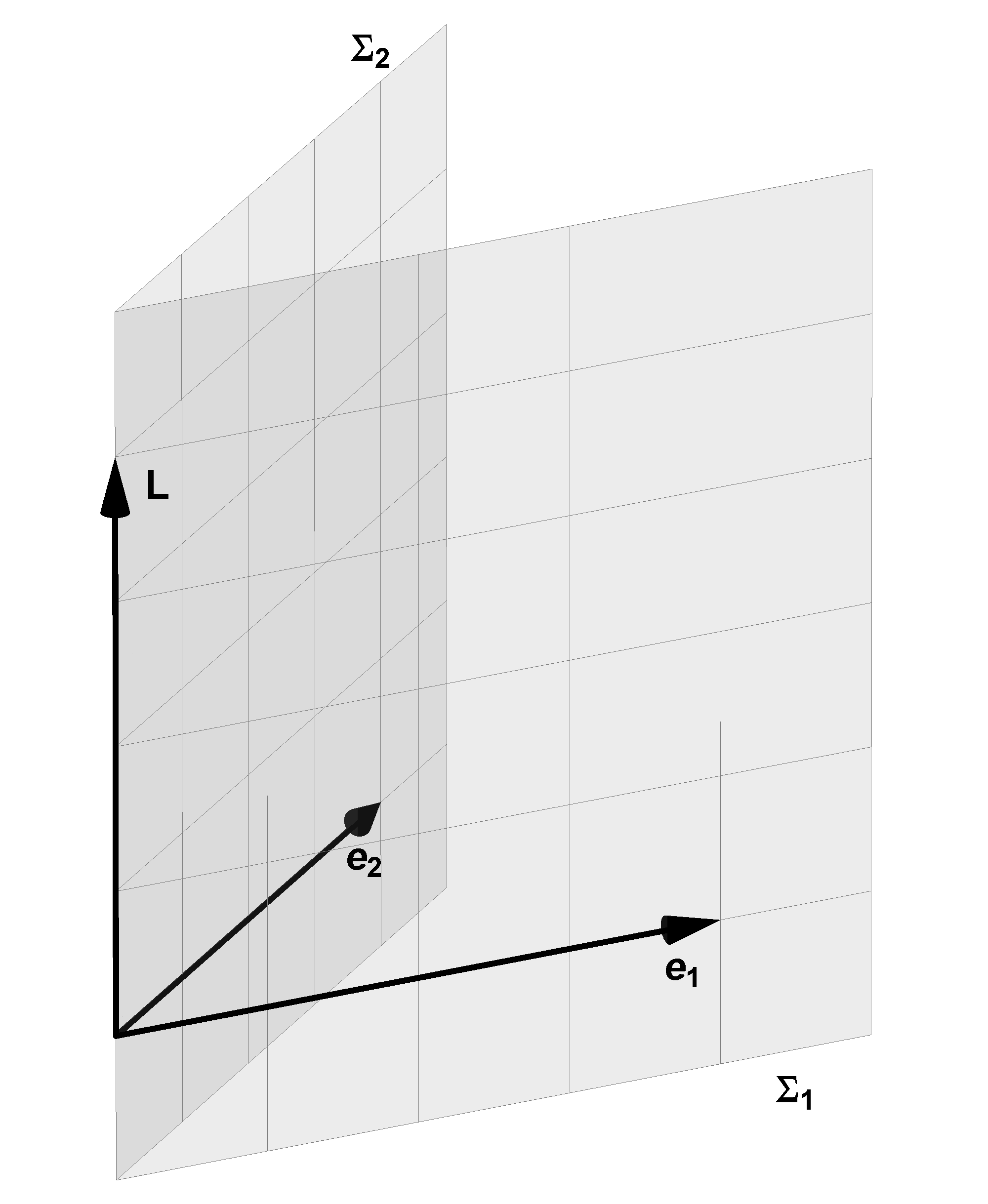}
\caption{The hyper-halfplanes $\Sigma_1$ and $\Sigma_2$ together form the boundary of a wedge of spacetime. If this wedge is removed and the opposite sides $\Sigma_1$ and $\Sigma_2$ are identified, then a conical singularity is formed along their common locus $L$. In the figure the time coordinate has been suppressed.}\label{fig:wedge}
\end{figure}

Given a holonomy one can always construct a conical defect with that holonomy by ``cutting'' away a wedge of spacetime and ``pasting'' the opposite sides together. The procedure is straightforward (see \cite{Krasnikov:2006zx} for a rigorous treatment). Consider a Lorentz transformation $Q$ that preserves a linear codimension 2 subspace $L$ of Minkoswki space $M$. Also pick a vector $e_1$ perpendicular to $L$. Since $Q$ is a Lorentz transformation the vector $e_2\equiv Q(e_1)$ is also perpendicular to $L$. The hyper-halfplanes $\Sigma_1$ and $\Sigma_2$, spanned by $L$ and respectively $e_1$ and $e_2$, form the opposite sides of a wedge in Minkoswki spacetime (See figure \ref{fig:wedge}).

By removing the interior of the wedge we obtain a spacetime with boundary (consisting of $\Sigma_1$ and $\Sigma_2$). The Lorentz transformation $Q$ provides a one-to-one mapping from $\Sigma_1$ to $\Sigma_2$, which we use to identify the points of $\Sigma_1$ with  the points of $\Sigma_2$. Similarly, the tangent spaces along $\Sigma_1$ and $\Sigma_2$ are identified using the map induced by $Q$ on the tangent space.

By construction, the result of this procedure is space that is locally isometric to Minkowski space, except along $L$. Moreover, it is straightforward to check that the holonomy of a simple loop around $L$ is given by $Q$. Also note, that the obtained geometry is independent of the choice of $e_1$.

In general, given an equal time slice $N$ of Minkoswski space, the Lorentz transformation $Q$ will not map the intersection of $\Sigma_1$ and $N$ to the intersection $\Sigma_2$ and $N$. Instead, the slice ``jumps'' across the wedge. However, there always exists a choice of $e_1$ such that $Q$ does map $\Sigma_1 \cap N$ to $\Sigma_2 \cap N$ \cite{hooft:1992}. This choice can be seen as setting $e_1$ and $e_2$ to be symmetric with respect to the direction of motion of $L$.
\begin{figure}[t]
\centering
\def\svgwidth{0.8\columnwidth}
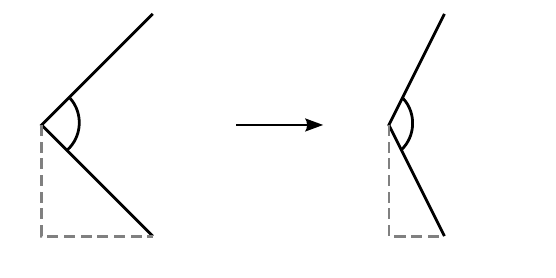
\caption{If a conical singularity singularity is boosted in the transverse direction, the deficit angle $\Delta\theta$ as measured by a stationary observer opens up. Shown projected in the $xy$-plane as seen by the stationary observer. }\label{fig:movingwedge}
\end{figure}
With this choice the geometry has well-defined equal time slices, and we can consider deficit angle as measured on such a slice. From figure \ref{fig:movingwedge}, we find that the deficit angle $\Delta\theta$ measured on a equal time slice for a moving string is
\begin{equation}
\Delta\theta=2\tan^{-1}(\frac{\tan\frac{\mu}{2}}{\cosh\chi}).
\end{equation}
Consequently, the deficit angle for the massless string in the Aichelburg--Sexl limit is
\begin{align}
\Delta\theta
&=\lim_{\chi\to\infty}2\tan^{-1}(\frac{\tan\frac{\bar\mu}{2\cosh\chi}}{\cosh\chi})\\
&=2\tan^{-1}\frac{\bar\mu}{2}.
\end{align}
This value is much more acceptable since it does not imply that the equal time slices become singular for finite values of $\bar\mu$.

We must note however, that there is not just a single choice for $e_1$ that will lead to a consistent equal time slicing. The vectors $e_1$ and $e_2$ just need to be symmetric with respect to the direction of motion of $L$. This means that the wedge can either point with or against the direction of motion. In fact, one could choose to split the wedge in two parts pointing forwards and backwards. As long as the sum of their parabolic angles is $\bar\mu$ this leads to the same spacetime geometry. However the measured deficit angle on an equal time slice is
\begin{equation}
\Delta\theta=2\tan^{-1}\frac{\bar\mu+\lambda}{2}+2\tan^{-1}\frac{\bar\mu-\lambda}{2},
\end{equation}
which can take any value between $0$ and $4\tan^{-1}\tfrac{\bar\mu}{2}$. We thus see that the deficit angle is still dependent on non-physical choices in the construction of the geometry.

\section{A smooth metric}\label{sec:smooth}
As we have seen, the metric \eqref{eq:masslessstring} for the massless cosmic string has some peculiar features along the $u=0$ hypersurface, even though the spacetime is completely flat everywhere except along the $y=0$ locus. One of these peculiar features is that geodesics crossing the $u=0$ hypersurface jump in the $v$ direction. This is easily fixed by shifting the $v$ coordinate in the opposite direction,
\begin{equation}
v \mapsto v + \frac{\bar\mu}{\sqrt{2}}\abs{y}\theta(u),
\end{equation}
where $\theta(u)$ is the unit step function. This results in the slightly better behaved metric
\begin{equation}\label{eq:masslessstring2}
\md s^2 =2\md u\md v + \md y^2 +\md z^2 +
\sqrt{2}\bar\mu\abs{y}\theta(u)\md u \md y.
\end{equation}
However, it is  discontinuous across the $y=0$ hypersurface  as well as the $u=0$ hypersurface, while only their common locus exhibits any curvature.

In this section we will construct a metric for the massless cosmic string that is completely smooth everywhere except along the worldsheet of the cosmic string. For this we first reconsider the metric for a stationary massive cosmic string
\begin{equation}\label{eq:massivestring3}
\md s^2 = -\md t^2 + \md r^2 + (1-\frac{\mu}{2\pi})^2 r^2\md\theta^2 +\md z^2.
\end{equation}
Part of the reason why this metric is so easy to express in cylindrical coordinates is that the holonomy of a conical defect acts simply as a shift of the $\theta$ coordinate. An even simpler way of representing the same geometry would be to take the Minkowski metric in cylindrical coordinates,
\begin{equation}
\md s^2 = -\md t^2 + \md r^2 +  r^2\md\theta^2 +\md z^2,
\end{equation}
and simply reduce the period of the $\theta$ coordinate by $\mu$. The factor $1-\frac{\mu}{2\pi}$ in \eqref{eq:massivestring3} serves to keep the period of $\theta$ equal to $2\pi$. In fact, the factor did not even need to be constant, any function of $\theta$ would serve the same purpose as long as its average over a period of $2\pi$ is equal to $1-\frac{\mu}{2\pi}$.

The idea in this section is to apply the same construction of rescaling an angular coordinate in ``cylindrical coordinates'' to a massless string. The first step is to find an analogue for ``cylindrical coordinates'' such that the holonomy of the string acts as a shift. For this we consider a general null-rotation $Q_\alpha$ that leaves the surface $u=y=0$ invariant. The transformation $Q_\alpha$ acts as follows on the lightcone coordinates
\begin{equation}
\begin{aligned}
Q(u) &= u\\
Q(v) &= v  -\tfrac{1}{2}\alpha^2 u+\alpha y\\
Q(y) &= y - \sqrt{2}\alpha u\\
Q(z) &= z.
\end{aligned}
\end{equation}
\begin{figure}[t]
\centering
\includegraphics[width=0.7\textwidth]{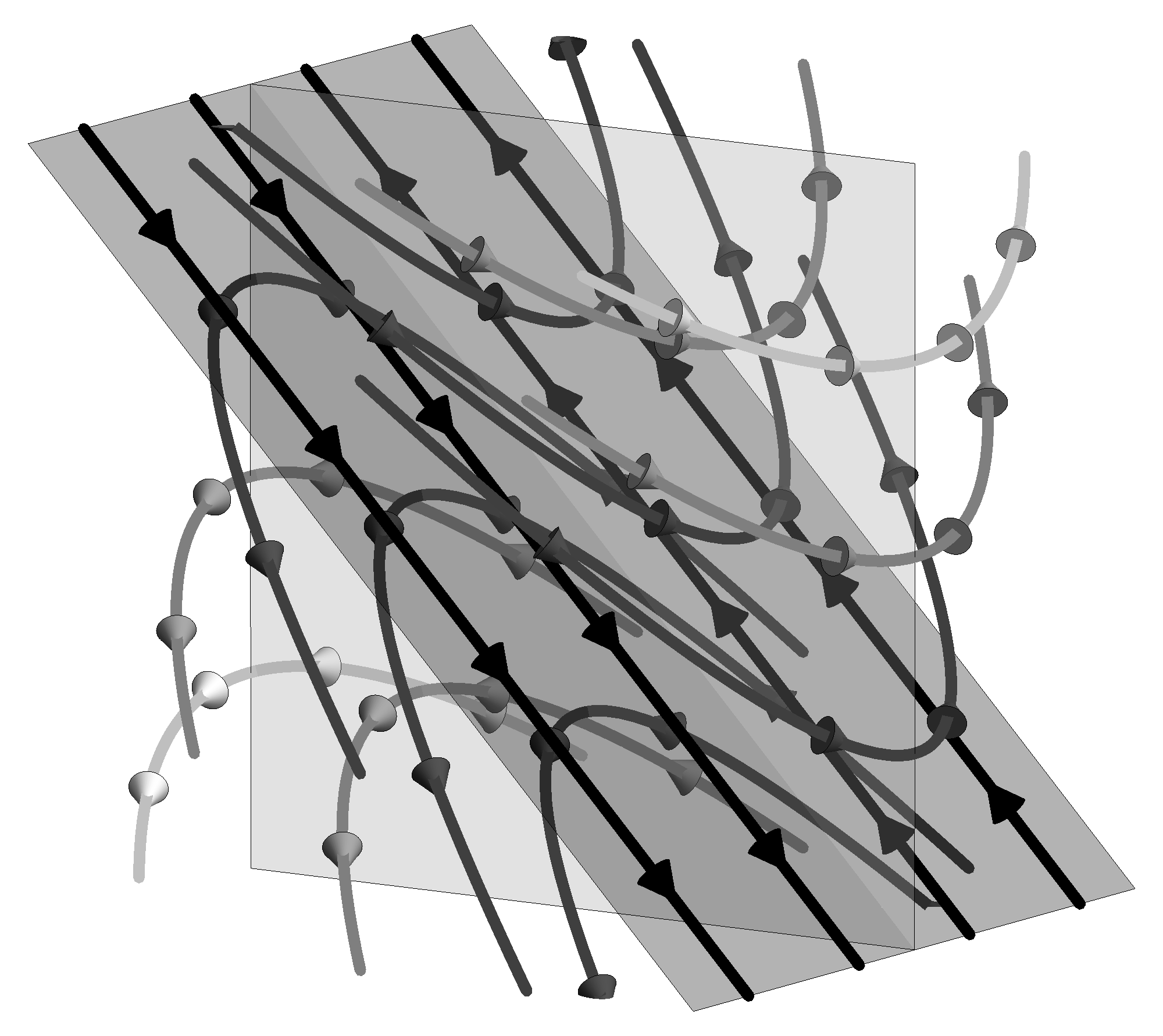}
\caption{The orbits of a null-rotation. Shown are the $u=0$ (dark) and $y=0$ (light) planes. The $z$-coordinate is suppressed.}\label{fig:orbits}
\end{figure}
The orbits of this action---shown in figure \ref{fig:orbits}---are parabolic curves parallel to the $u=0$ plane. To achieve our objective that the holonomy acts as a shift on the coordinates we want the orbits of $Q_\alpha$ to coincide with the constant coordinate lines, i.e. each orbit needs to be assigned three coordinates.
For the first three coordinates ($\bar u$, $\bar v$, $\bar z$) we choose the values of $u$, $v$ and $z$ where each orbit intersects with the $y=0$ plane. As the fourth coordinate of a point we take the parabolic angle $\alpha$ by which it is removed from this plane. This yields the following set of coordinates
\begin{equation}
\begin{aligned}
 \bar{u} &= u\\
 \bar{v} &= v + \tfrac{1}{2}y^2/u \\
 \alpha &= y/u\\
 \bar{z} &= z.
\end{aligned} 
\end{equation}
In these coordinates the Minkowski metric becomes
\begin{equation}\label{eq:nullcoorinates}
\md s^2 = 2\md\bar u\md\bar v + \bar{u}^2\md \alpha^2  +\md z^2,
\end{equation}
which is similar to the Minkowski metric in cylindrical or Rindler coordinates except that the ``radial'' coordinate $\bar{u}$ is  lightlike rather than spacelike or timelike respectively. Like those coordinate systems, it is singular when the radial coordinate vanishes. In this case this is the $\bar u=0$ surface, which cuts the space in two coordinate patches. 

However, unlike the angular coordinate in cylindrical coordinates, the range of the angular coordinate $\alpha$ is infinite. Therefore, a rescaling of the angular coordinate does not lead to the desired deficit angle. Instead, we replace the $\bar u >0$ coordinate patch of the metric \eqref{eq:nullcoorinates} by
\begin{equation}\label{eq:smoothmetric1}
\md s^2 = 2\md\bar u\md\bar v + \bhh{1-f(\alpha)}^2\bar{u}^2\md \alpha^2  +\md z^2,
\end{equation}
where $f(\alpha)$ is a smooth function with compact support satisfying $f(\alpha)<1$. It is straightforward to verify that this metric is indeed Riemann flat on the entire coordinate patch, and because $f(\alpha)$ has compact support it seamlessly connects to the $\bar u <0$ patch across the entire  original $u=0$ hyperplane excluding the original $y=0$ axis. Consequently, the newly constructed spacetime is Riemann flat everywhere except possibly the original $u=y=0$ surface.

We can thus conclude that the newly constructed metric is locally equivalent the metric for the massless cosmic string \eqref{eq:masslessstring} everywhere away from the null surface $u=y=0$. To verify that they are fully equivalent we need to match the holonomy of a loop around this null surface to the holonomy of the conical defect. Since the new metric is flat everywhere, we are at liberty to choose the path to suit the calculation. We first choose two values $\alpha_1 < \alpha_2$ such that $f(\alpha)=0$ outside the interval $(\alpha_1,\alpha_2)$. As one arc of the loop we choose
\begin{equation}
\gamma^\mu(\lambda) = (\frac{1}{1-f(\lambda)},0,\lambda,0),
\end{equation}
with $\lambda$ running from $\alpha_1$ to $\alpha_2$.

On this arc, the integrand of holonomy integral is
\begin{equation}
-\Gamma^{\mu}_{\sigma\nu}\frac{\md \gamma^\sigma}{\md \lambda}=\begin{pmatrix}
0 & 0 & 0 & 0\\
0 & 0 & f(\lambda)-1  & 0\\
1-f(\lambda) &0 & 0 & 0\\
0 & 0 & 0 & 0
\end{pmatrix}.
\end{equation}
Since the integrand commutes with itself for all values of $\lambda$ the path ordering is trivial and we find for the parallel propagator

\begin{equation}
(P_\gamma)_{\phantom{\mu}\nu}^\mu =\begin{pmatrix}
1 & 0	 & 0 & 0\\
-\frac{1}{2}\bhh{\alpha_2-\alpha_1-I}^2
 	& 1
		& I - \alpha_2 + \alpha_1
			& 0\\
\alpha_2-\alpha_1-I
	& 0
		& 1
			& 0\\
0 & 0 & 0 & 1
\end{pmatrix},
\end{equation}
where
\begin{equation}
I\equiv \int_{\alpha_1}^{\alpha_2} f(\alpha)\md\alpha =\int_{-\infty}^{\infty} f(\alpha)\md\alpha.
\end{equation}   
Outside the region $\bar u>0$ and $\alpha_1<\alpha<\alpha_2$ metric \eqref{eq:smoothmetric1} is identical to the metric \eqref{eq:nullcoorinates}. Since \eqref{eq:nullcoorinates} is just the Minkowski metric, the holonomy of any loop must equal the identity. Consequently, any closing arc (as long as it stays out of the $\bar u>0$ and $\alpha_1<\alpha<\alpha_2$ region) has a parallel propagator equal to the inverse of the parallel propagator along $\gamma$ with $f(\alpha)$ set to zero.
 
As a result, the holonomy of a closed loop is equal to 
\begin{equation}
Q_{\phantom{\mu}\nu}^\mu =\begin{pmatrix}
1 & 0	 & 0 & 0\\
-\frac{1}{2}I^2
 	& 1
		& I
			& 0\\
-I
	& 0
		& 1
			& 0\\
0 & 0 & 0 & 1
\end{pmatrix}.
\end{equation}
This is indeed a null-rotation of the form \eqref{eq:Hmstringlc}. Consequently, the two holonomies match if we choose $f(\alpha)$ such that $I=\sqrt{2}\bar\mu$. Since the spacetimes \eqref{eq:masslessstring} and \eqref{eq:smoothmetric1} are Riemann flat everywhere away from the $u=y=0$ surface, this implies that the holonomies of \emph{all} possible loops match.  This is enough to conclude that \eqref{eq:masslessstring} and \eqref{eq:smoothmetric1} are equivalent. 

However, this metric is still singular along the $\bar u =0$ plane. To obtain a metric that is manifestly smooth and non-singular everywhere except at the conical singularity, we change coordinates back to normal lightcone coordinates in which the constructed metric becomes
\begin{equation}
\md s^2 = 2\md u\md v + \md y^2  +\md z^2 - \bhh{2-f(\tfrac{y}{u})}f(\tfrac{y}{u})\theta(u)\frac{(u\md{y}-y\md{u})^2}{u^2},
\end{equation}
where $\theta(u)$ is the unit step function. This metric is: a) smooth everywhere except along the $u=y=0$ locus, b) equivalent to the massless string metric \eqref{eq:masslessstring}. The downside is that this metric is no longer equal to the Minkowski metric almost everywhere.

\section{Discussion}
We have identified the geometry generated by a massless cosmic string as a flat spacetime with a conical defect which has a null-rotation as its holonomy. Since all null-rotations belong to the same conjugation class of the Lorentz group, this is consistent with the observation in \cite{PHDthesis} that the mass (density) of a conical defect can be identified with the conjugation class of its holonomy. Since the surrounding spacetime is Riemann flat everywhere, there is no accompanying gravitational shockwave as had been previously suggested in the literature based on the $\delta$-singularity in the metric obtained through the Aichelburg--Sexl limiting procedure. To further illustrate that the $u=0$ surface is unremarkable, we explicitly constructed a metric for the massless cosmic string which is smooth everywhere in its surroundings.

In this paper we have not touched on the subject of how massless cosmic strings could be formed in nature. The author's interest in the subject originated in 't Hooft's locally finite piecewise flat model for gravity in 3+1 dimensions, where massless cosmic strings arise naturally as fundamental excitations of the model. However, it is more conventional to study cosmic strings formed in symmetry breaking phase transitions in the early universe. The mass of such cosmic strings is determined by a topological winding number, implying that there is only a discrete set of possible masses. Consequently, the massless limit as considered in this paper does not exist. Nonetheless, the description given here may be useful as an approximation of ultra-relativistic traditional cosmic strings.

Even if it were possible to form massless cosmic strings as the result of a gauge theoretic phase transition (and the author is not aware of any process that would allow this), it remains unclear if the resulting configuration would be stable. The normal stabilizing factor provided by the topological charge is likely to be absent. Moreover, as can be seen from equation \eqref{eq:EMmstring}, the massless cosmic strings discussed here have zero tension in the direction of the string, indicating that there is essentially nothing holding the string together and casting further doubt on their stability. Perhaps the best bet for a realization of massless cosmic strings would be as massless modes of macroscopic fundamental strings. Although the author is again not aware of any specific model that would allow this.

\section*{Acknowledgments}
The author gratefully acknowledges Gerard 't Hooft for the support during the preparation of this paper.
\bibliographystyle{apsrev4-1}
\bibliography{journalshortnames,mstring}
\end{document}